\documentclass[fleqn,usenatbib]{mnras}

\usepackage{newtxtext,newtxmath}

\usepackage[T1]{fontenc}

\DeclareRobustCommand{\VAN}[3]{#2}
\let\VANthebibliography\thebibliography
\def\thebibliography{\DeclareRobustCommand{\VAN}[3]{##3}\VANthebibliography}

\usepackage{graphicx}	
\usepackage{amsmath}	

\newcommand{\jwst}{{\em JWST}}
\newcommand{\hst}{{\em HST}}
\newcommand{\muv}{$M_{\rm UV}$}

\title[The Ultraviolet Luminosity Function at $z>9$]{JWST UNCOVER: The Overabundance of Ultraviolet-luminous Galaxies at $z>9$}

\author[Chemerynska et al.]{Iryna Chemerynska$^{1}$
\thanks{E-mail: iryna.chemerynska@iap.fr},
Hakim Atek$^{1}$,
Lukas J. Furtak$^{2}$,
Adi Zitrin$^{2}$,
Jenny E. Greene$^{3}$,
Pratika Dayal$^{4}$,
\newauthor Andrea Weibel$^{5}$,
Seiji Fujimoto$^{6}$,
Vasily Kokorev$^{4}$,
Andy D. Goulding$^{7}$,
Christina C. Williams$^{8,9}$,
\newauthor Themiya Nanayakkara$^{10,11}$,
Rachel Bezanson$^{12}$,
Gabriel Brammer$^{13}$,
Sam E. Cutler$^{14}$,
Ivo Labbe$^{9}$,
\newauthor Joel Leja$^{15,16,17}$,
Richard Pan$^{18}$,
Sedona H. Price$^{12}$,
Pieter van Dokkum$^{19}$,
Bingjie Wang$^{15,16,17}$,
\newauthor John R. Weaver$^{14}$,
Katherine E. Whitaker$^{13,14}$
\\
$^{1}$Institut d'Astrophysique de Paris, CNRS, Sorbonne Universit\'e, 98bis Boulevard Arago, 75014, Paris, France\\
$^{2}$Physics Department, Ben-Gurion University of the Negev, PO Box 653, Be’er-Sheva 8410501, Israel\\
$^{3}$Department of Astrophysical Sciences, Princeton University, 4 Ivy Lane, Princeton, NJ 08544\\
$^{4}$Kapteyn Astronomical Institute, University of Groningen, P.O. Box 800, 9700 AV Groningen, The Netherlands\\
$^{5}$Department of Astronomy, University of Geneva, Chemin Pegasi 51, 1290 Versoix, Switzerland\\
$^{6}$Department of Astronomy, The University of Texas at Austin, Austin, TX 78712, USA\\
$^{7}$Department of Astrophysical Sciences, Princeton University, Princeton, NJ 08544, USA\\
$^{8}$NSF's National Optical-Infrared Astronomy Research Laboratory, Tucson, AZ 85719, USA\\
$^{9}$Steward Observatory, University of Arizona, Tucson, AZ 85721, USA\\
$^{10}$Centre for Astrophysics and Supercomputing, Swinburne University of Technology, Melbourne, VIC 3122, Australia\\
$^{11}$ARC Centre of Excellence for All Sky Astrophysics in 3 Dimensions (ASTRO-3D), Australia\\
$^{12}$Department of Physics and Astronomy and PITT PACC, University of Pittsburgh, Pittsburgh, PA 15260, USA\\
$^{13}$Cosmic Dawn Center (DAWN), Niels Bohr Institute, University of Copenhagen, Jagtvej 128, K{\o}benhavn N, DK-2200, Denmark\\
$^{14}$Department of Astronomy, University of Massachusetts, Amherst, MA 01003, USA\\
$^{15}$Department of Astronomy \& Astrophysics, The Pennsylvania State University, University Park, PA 16802, USA\\
$^{16}$Institute for Computational \& Data Sciences, The Pennsylvania State University, University Park, PA 16802, USA\\
$^{17}$Institute for Gravitation and the Cosmos, The Pennsylvania State University, University Park, PA 16802, USA\\
$^{18}$Department of Physics \& Astronomy, Tufts University, MA 02155, USA\\
$^{19}$Astronomy Department, Yale University, 219 Prospect St, New Haven, CT 06511, USA
}

\date{Accepted XXX. Received YYY; in original form ZZZ}

\pubyear{2023}

\begin{document}
\label{firstpage}
\pagerange{\pageref{firstpage}--\pageref{lastpage}}
\maketitle

\begin{abstract} 
Over the past year, JWST has uncovered galaxies at record-breaking distances up to $z \sim 13$. The JWST UNCOVER (ultra-deep NIRSpec and NIRcam observations before the epoch of reionization) program has obtained ultra-deep multiwavelength NIRCam imaging of the massive galaxy cluster Abell 2744 over $\sim 45$ arcmin$^{2}$ down to $\sim 29.5$ AB mag. Here, we present a robust 
ultraviolet (UV) luminosity function derived through lensing clusters at $9<z<12$. Using comprehensive end-to-end simulations, we account for all lensing effects and systematic uncertainties in deriving both the amplification factors and the effective survey volume. Our results confirm the intriguing excess of UV-bright galaxies (\muv\ $< -20$ AB mag) previously reported at $z>9$ in recent JWST studies. In particular, a double power-law (DPL) describes better the bright-end of the luminosity function compared to the classical Schechter form. The number density of these bright galaxies is 10-100 times larger than theoretical predictions and previous findings based on Hubble Space Telescope (HST) observations. Additionally, we measure a star formation rate density of $\rho_{\rm SFR} = 10^{-2.64}$ M$_{\odot}$~yr$^{-1}$~Mpc$^{-3}$ at these redshifts, which is 4 to 10 times higher than galaxy formation models that assume a constant star formation efficiency. Future wide-area surveys and accurate modeling of lensing-assisted observations will reliably constrain both the bright and the dim end of the UV luminosity function at $z>9$, which will provide key benchmarks for galaxy formation models. 
\end{abstract}




\section{Introduction}\

The advent of the {\em James Webb Space Telescope} (\jwst) has opened a new era in our exploration of the distant universe. While the {\em Hubble Space Telescope} (\hst) helped identify more than 2,000 dropout galaxy candidates at $z>6$ \citep{bouwens21}, its restricted near-infrared coverage, and the limited capabilities of the {\em Spitzer Space Telescope} hampered the identification of galaxies at $z>9$ and the characterization of their stellar populations. The unprecedented capabilities of JWST's Near Infrared Camera \citep[NIRCam;][]{rieke23}, which uniformly covers the $0.8-5\mu m$ spectral range with exquisite spatial resolution, provides the opportunity to routinely identify $z>9$ galaxies. 

The first \jwst\ observing programs, which included the Early Release Observations (ERO) and the Early Release Science Programs (ERS) \citep[e.g.,][]{pontoppidan22,treu22,finkelstein22}, have unveiled a large sample of galaxies at redshifts higher than $z=9$ \citep{atek23,castellano23,donnan23,finkelstein22b,morishita22,naidu22,robertson22,bunker23,adams23,austin23}. These sources have been photometrically identified via their Lyman-break signature, photometric redshifts derived from model fitting to their spectral energy distribution (SED), or a combination thereof. Furthermore, spectroscopic follow-up campaigns with the Near-infrared Spectrograph \citep[NIRSpec;][]{jakobsen22} instrument has allowed us to confirm the high-redshift nature of these candidates at an unprecedented rate \citep[e.g.][]{curtis-lake23,arrabal23,hainline2023,roberts-borsani22,roberts-borsani23,hsiao23,bunker23,wang23,fujimoto23}.

One of the most prominent results is the surprising over-abundant population of UV-bright galaxies at $z\sim9-12$ when compared to theoretical predictions of the UV luminosity function at those redshifts \citep[e.g.][]{naidu22, harikane23, bouwens23, mauerhofer23}. Indeed, most models suggest that galaxy evolution should extend beyond $z>10$ \citep[e.g.;][]{hutter21,dayal22}, including a rapid decline in the star formation efficiency. However, strong observational constraints have been lacking until the recent \jwst\ results. Given the small area probed by these surveys, the expected rapid evolution predicts more than an order of magnitude fewer galaxies than observed. This small survey area could also result in a significant cosmic variance \citep[e.g.;][]{ucci21}. From the theoretical point of view, several physical explanations have been investigated, which are mostly based on the star formation histories and the stellar population properties of these galaxies. Some suggest that the expected decline in the number density of galaxies with redshift is balanced by the decrease of dust content \citep{ferrara22}, while others suggest the inefficient star-formation feedback \citep{dekel23} or a weaker pre-reionization background \citep{harikane23} result in a larger population of bright galaxies at $z>10$. According to \cite{parashari23}, an increase in the primordial power spectrum can explain the high stellar masses in high-redshift massive galaxies, assuming low to moderate star formation efficiency. Stochastic star formation histories (SFH) have also been invoked to explain this excess, resulting in notable disagreements across the literature. Some show that bursty SFHs cause UV luminosity deviations large enough to explain these observations \citep{shen23,sun2023,munoz23,whitler23} while others find it to be insufficient \citep{pallottini23,ciesla23,mason23}. An evolution in the initial mass function
(IMF) or UV contribution from active galactic nuclei \citep[AGN][]{ono18,fujimoto23} can also explain such UV excess \citep{pacucci22}. Indeed, the past year of \jwst\ observations has revealed an astonishing number of optically red AGN at $z>6$ that have been entirely missed by previous UV-based selections \citep{matthee23,furtak23b,furtak23c,labbe23b,greene23,maiolino23}. A few galaxy candidates have shown signs of black hole activity via their emission lines or their X-ray emission \citep{kokorev23,goulding23,fujimoto23,larson23}, potentially proving a non-negligible contribution to the bright-end of the UV luminosity function. While their number density is still an order of magnitude lower than that of galaxies, it appears to be larger than previous UV-based determinations at $z=6$. Another potential source of contamination is the low-redshift dusty galaxies that can mimic the broadband colors of high-redshift galaxies \cite[e.g.;][]{naidu22b,zavala23,arrabal23}.

In this work, we present the galaxy UV luminosity function (UV LF) over the redshift range $z=9-12$ based on the lensing-assisted observations from the \jwst\ UNCOVER program (PIs: Labb\'e \& Bezanson,~JWST-GO-2561). The observations consist of deep NIRCam imaging of the Abell 2744 cluster (A2744), which is at the redshift $z = 0.31$. This work builds on a comprehensive assessment of the lensing effects impacting the UV LF using end-to-end simulations from the source plane to the final UV LF. The paper is organized as follows: in Section \ref{sec:Observations} we describe the imaging data-set used in the study and the lensing model is covered in Section \ref{sec:Lensing}.  We present the high-redshift galaxies catalog at $z > 9$ and mock galaxies at the same redshift in Section \ref{sec:High-z}. Our forward modeling procedure and the final UV LF with the associated uncertainties are presented in Section \ref{sec:UVLF} while the star formation rate density in the early Universe is detailed in Section \ref{sec:sfrd}. The conclusion is given in Section \ref{sec:Conclutions}. 

Throughout this work, we assume a flat $\Lambda$CDM cosmology with $H_0$ = 70 km s$^{-1}$ Mpc$^{-1}$, $\Omega_{M}$ = 0.3 and $\Omega_{\Lambda}$ = 0.7.

\section{Observations}\
\label{sec:Observations}
The first part of UNCOVER consisted of multi-wavelength NIRCam imaging of the lensing cluster Abell 2744 in 6 broadband filters (F115W, F150W, F200W, F277W, F356W, and F444W) and one medium-band filter (F410M), reaching a limiting magnitude of $\sim 29.5$ AB over $\sim 45$ arcmin$^{2}$ \citep{bezanson22}. 
Observations were obtained across multiple days, between November 2nd and 15th 2022. The \jwst\ imaging data also include parallel observations with the Near Infrared Imager and Slitless Spectrograph (NIRISS), using five broadband filters — F115W, F150W, F200W, F356W, and F444W. We also incorporate imaging data obtained by the ERS program GLASS \citep[PI: Treu, JWST-DD-ERS-1324,][]{treu22} and the DDT (Director Discretionary Time) program ID 2756 (PI: Chen, JWST-DD-2756), which cover the outer regions of the UNCOVER field and which use a nearly identical filter set, except for the addition of F090W for the former, and the absence of F444W for the latter, respectively.   

\begin{table}
    \centering
\caption{UNCOVER Imaging: Limiting AB magnitudes at 5$\sigma$, measured in 0.32" diameter apertures, correspond to the area-weighted 90th percentiles \citep{weaver23}.
}
\label{tab:UNCOVER_Imaging}
    \begin{tabular}{cccc}  \hline \hline 
 Filter& Exposure&Depth (AB)& Area (arcmin$^{2}$)\\ \hline \hline 
         F435W& 12.5h &28.30&15.89\\
         F606W& 7.0h &27.42&33.11\\
         F814W& 29.0h &27.17&28.05\\
         F090W& 5.3h & 29.30&11.65\\ 
         F105W& 16.7h &27.06&17.95\\
         F115W& 6.0h & 28.15&44.53\\  
         F125W& 8.3h &27.09&17.48\\
         F140W& 7.0h &28.54&4.84\\
         F150W& 6.0h & 28.23&45.00\\  
         F160W& 16.7h &26.53&18.89\\
         F200W& 3.7h & 28.47&44.10\\  
         F277W& 3.7h & 28.76&44.11\\  
         F356W& 3.7h & 28.88&44.20\\  
         F410M& 3.7h & 28.58&25.90\\ 
         F444W& 4.6h & 28.24&44.37\\ \hline
    \end{tabular}
\end{table}

The reduction of the imaging data was performed using the {\tt grizli} \citep{brammer22} software and presented in \citet{weaver23}. 
Flux-calibrated NIRCam exposures from Stage 2b of the JWST calibration pipeline (v1.8.4) are combined with calibration set {\tt jwst\_1039.pmap} to create imaging mosaics. The GRIsm redshift and LIne analysis software for space-based spectroscopy was used to process, align and coadd the exposures \citep[GRIzLI,1.8.16.dev12;][]{brammer19, Kokorev22}. The data were drizzled into mosaics with a pixel scale of 0.04$\arcsec$ pix$^{-1}$ \citep{Atek_2023b}. The exposure times with the resulting magnitude limits in all filters are listed in Table \ref{tab:UNCOVER_Imaging}. 

The galaxy cluster A2744 was part of the Hubble Frontier Fields (HFF) clusters. As such, it has deep {\em Hubble Space Telescope} (\hst) imaging data across the optical and near-infrared range, including ancillary data from multiple programs \citep{lotz17}. These data include deep Advanced Camera for Surveys (ACS) imaging in the cluster core (F435W, F606W, and F814W) and Wide-Field Camera Three (WFC3) observations in 4 filters (F105W, F125W, F140W, and F160W). In addition, A2744 was observed as part of the \hst\ program BUFFALO \citep{steinhardt20}. The ACS observations included three broadband filters (F435W, F606W, F814W) while the WFC3 used four filters (F105W, F125W, F140W, F160W). The \jwst\ and \hst\ observations were drizzled to the same pixel scale and aligned. \cite{weaver23} contains a thorough discussion of the data reduction procedure. 

\section{Gravitational lensing}\
\label{sec:Lensing}
To compute the UV LF at $z$ = 9 - 12, we need to have a good understanding of the lensing power of the galaxy cluster A2744. Lensing models are crucial not only for estimating the magnification of the sources but also for estimating the effective survey volume. In this work, we adopt the \texttt{v1.1} UNCOVER strong lensing (SL) model of A2744, presented in \cite{furtak23a}, which is publicly available on the UNCOVER website\footnote{\url{https://jwst-uncover.github.io/DR2.html\#LensingMaps}}. This model was constructed using a new version of the parametric code by \cite{zitrin2015}, updated to be fully analytic and thus not dependent on a fixed grid, which allows for faster computation and with a higher resolution \citep{pascale2022,furtak23a}. The model for A2744 comprises five smooth cluster-scale dark matter halos, centered on the five brightest cluster galaxies (BCGs), and 421 cluster member galaxies, as detailed in \citep{furtak23a}. More than half of the cluster galaxies are spectroscopically confirmed \citep{bergamini23a}. The \texttt{v1.1} of the model used here is constrained by a total of 141 multiple images (belonging to 48 sources), of which 96 have spectroscopic redshifts \citep{bergamini23a,bergamini23b,roberts-borsani23} and the remaining ones are photometric systems discovered with the UNCOVER imaging \citep{furtak23a,furtak23b}. With these constraints, the model achieves a lens plane image reproduction RMS of $\Delta_{\mathrm{RMS}}=0.51\arcsec$. We compute analytic magnifications and their uncertainties for our sample at each object's position and spectroscopic redshift. 


\begin{figure*}
    \centering
    \includegraphics[width = 1.0 \textwidth]{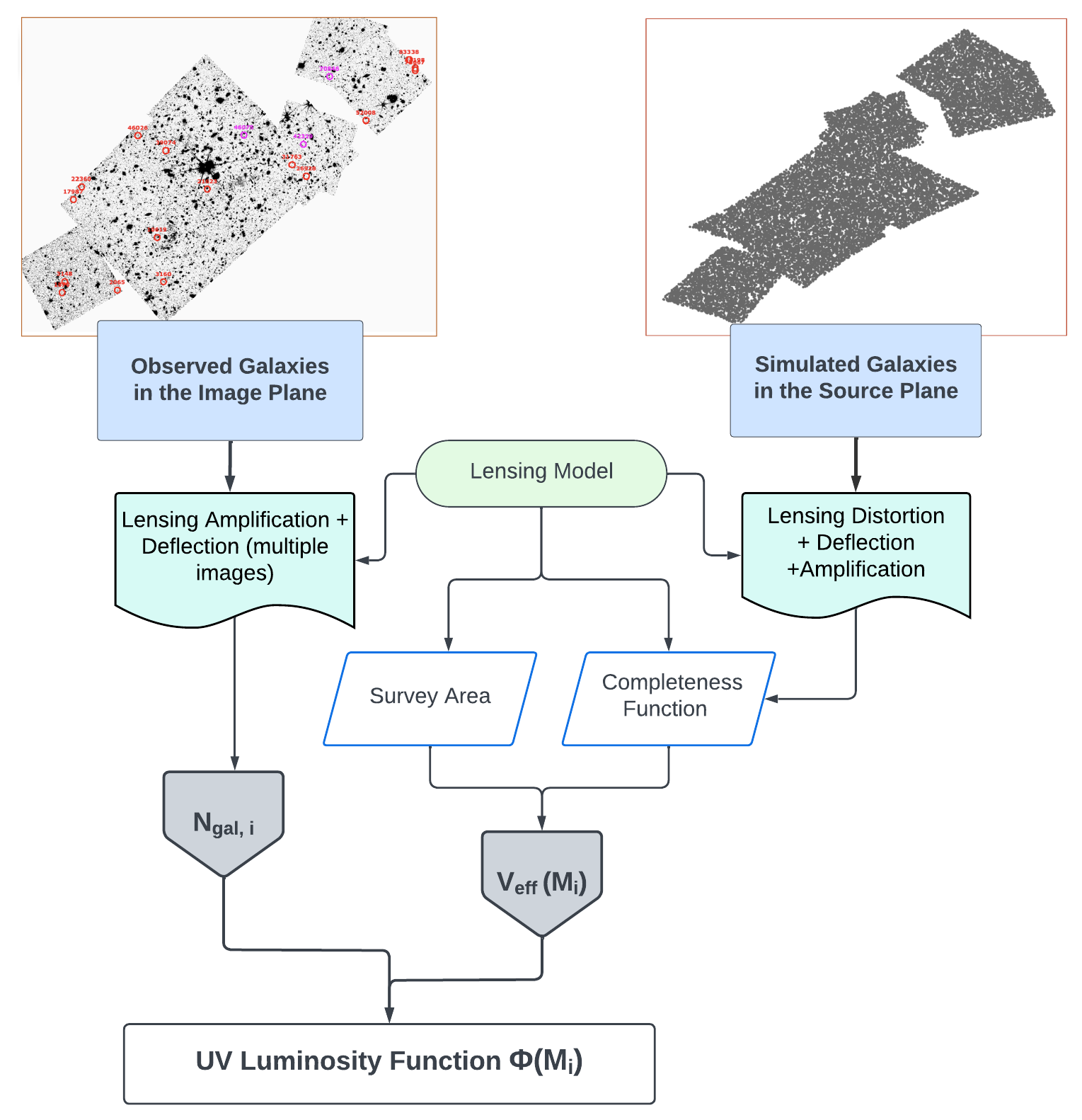}
    \caption{This schematic illustrates the process of computing the UV LF. The first part is the detection of lensed galaxies in the observed image and estimates of their lensing amplification and multiplicity to determine the observed number counts. The second step is the survey completeness, which consists of simulating galaxies directly in the source plane before projecting them back to the image plane using the same lensing model used for observations. The completeness function, informed by simulated sources, is combined with survey area reduction as a function of magnification to determine the effective survey volume. The original footprint, which includes UNCOVER, GLASS, DDT observations, can be found in \citep{Atek_2023b}.}
    \label{fig:LF_computation}
\end{figure*}

\section{High-redshift galaxy sample}
\label{sec:High-z}

\subsection{Observed galaxies}\
The first step is to select a sample of dropout galaxies in the imaging data of A2744, before proceeding to compute their number density. In this work, we rely on the sample selection and the derived high-redshift galaxy catalog presented in \cite{Atek_2023b}.
The general UNCOVER catalog is presented in \cite{weaver23}. However, our high-redshift selection is based on a second photometric catalog, which is tailored towards high-redshift sources, and presented in \citet{Atek_2023b}. The full description of this catalog will be given in Weibel et al. (in prep). Briefly, this catalog was produced by using the {\tt SExtractor} tool \citep{bertin96} in its dual mode, using the F444W as the detection image. Before performing the aperture photometry, we matched the point spread function (PSF) of each image to the longest wavelength image in the F444W filter. Instead of using simulated PSFs from WebbPSF \citep{perrin14}, we built empirical ones from the NIRCam data, by following the procedure described in \cite{skelton2014} and \cite{whitaker2019}. 
Fluxes in each filter are measured in 0.24\arcsec\ apertures, whereas total fluxes are derived by applying a scaling factor calculated from the ratio of aperture flux and F444W {\tt SExtractor AUTO FLUX}.  We adopted the Lyman break selection criteria defined in \cite{atek23} to define the selection window.
Galaxies at redshift from 9 to 11 need to satisfy the following color-color criteria:
\begin{equation}
	\begin{array}{l}
		M_{115}-M_{150}>1.0\\
		M_{115}-M_{150}>1.5+1.4(M_{150}-M_{200})\\
		M_{150}-M_{200}<0.5
	\end{array}
  \label{eq:z9}
\end{equation}
For $12<z<15$ galaxies we adopt the following criteria:
\begin{equation}
	\begin{array}{l}
		M_{150}-M_{200}>1.5\\
		M_{150}-M_{200}>1.6+1.8(M_{200}-M_{277})\\
		M_{200}-M_{277}<0.5
	\end{array}
 \label{eq:z11}
\end{equation}

This color selection was done to minimize the contamination rate. The most important sources of contamination consist of dust-obscured and evolved galaxies at lower redshift with extremely red colors, and low-mass stars. In addition to selection criteria, we also require that sources are detected in all LW filters with a minimum SNR = 5 and that they remain undetected in F090W, when available, and all HST optical bands at a 2$\sigma$ level. To ensure a minimum continuum break of one magnitude, we assign a 1$\sigma$ lower limit for sources not detected in the dropout filter, corresponding to the filter limiting magnitude. All details are described in \cite{Atek_2023b}.

\begin{table*}
\centering
\caption{The photometric and physical characteristics of the sample of high-redshift candidates identified through the Abell 2744 cluster. The object presented with an asterisk indicates the potential AGN source in our sample. Source IDs correspond to those used in \citet{Atek_2023b} and ID(W23) used in the UNCOVER photometric catalog of \citet{weaver23}. For the spectroscopically-confirmed sources quoted in \citet{fujimoto23}, we also provide their confirmed redshift. }
\begin{tabular}{l|c|c|c|c|c|c|c|c|c}
\hline
\hline
ID & ID(W23) &RA & Dec & $z_{\mathrm{phot}}$ &$z_{\mathrm{spec}}$ & $m_{\mathrm{F444W}}$&$M_{\mathrm{UV}}$ &  $\mu$ \\\hline
\hline
1870 & 3342 &3.648010 &-30.426616 & $9.32_{-6.95}^{+0.96}$ & & 26.18&-19.78 $\pm$ 0.18 & $1.30_{-0.01}^{+0.01}$\\
3148&4808  &3.646481 &-30.421615 & $9.40_{-7.14}^{+0.88}$ &&25.01&-20.51 $\pm$ 0.18 & $1.31_{-0.01}^{+0.02}$\\
17987 &22335 &3.641572 &-30.382825 & $9.41_{-7.12}^{+0.60}$ &&26.2&-19.39 $\pm$ 0.18 &  $1.31_{-0.01}^{+0.01}$\\
26928 &29903  &3.511925 &-30.371861 & $9.47_{-0.07}^{+0.44}$ &&24.81&-20.36 $\pm$ 0.12 &  $1.67_{-0.09}^{+0.07}$\\
2065 &3686  &3.617194 &-30.425536 & $9.50_{-0.08}^{+0.34}$ &$9.325_{-0.001}^{+0.000}$ &23.72 &-21.67 $\pm$ 0.12 & $1.65_{-0.03}^{+0.02}$\\
83338& 55807 &3.454706 &-30.316898 & $9.55_{-0.57}^{+0.91}$ &&25.79&-19.24 $\pm$ 0.18 &  $1.17_{-0.01}^{+0.01}$\\ 
10619&13935 & 3.594996 &-30.400738 & $9.69_{-0.12}^{+0.33}$ &&26..41&-17.57 $\pm$ 0.13 &   $11.50_{-0.50}^{+0.40}$\\
21623* &26185*  &3.567067 &-30.377869 & $10.01_{-0.26}^{+0.36}$ &$10.071_{-0.001}^{+0.000}$ &25.81&-19.01 $\pm$ 0.14 &  $3.72_{-0.18}^{+0.14}$\\
52008&44832  &3.478739 &-30.345535 & $10.37_{-1.09}^{+0.32}$ &&26.03&-19.90 $\pm$ 0.14 &  $1.26_{-0.02}^{+0.02}$\\
81198&54706  &3.451367 &-30.320717 & $10.50_{-0.66}^{+0.23}$ &&26.47&-19.90 $\pm$ 0.14 &  $1.17_{-0.01}^{+0.01}$\\
39074 &37126  &3.590115 &-30.359743 & $10.60_{-0.31}^{+0.21}$ &$10.255_{-0.001}^{+0.001}$&25.89 &-20.03 $\pm$ 0.14 & $1.89_{-0.06}^{+0.05}$\\
73667 &54328 &3.451412 &-30.321807 & $10.68_{-0.31}^{+0.21}$ &&26.26&-20.55 $\pm$ 0.13 &  $1.17_{-0.01}^{+0.01}$\\
22360&26136  &3.637111 &-30.376780 & $10.73_{-1.19}^{+0.44}$ &&25.58&-19.85 $\pm$ 0.18 &  $1.33_{-0.01}^{+0.01}$\\
3160 &4890 &3.591436 &-30.421663 & $10.74_{-1.45}^{+0.37}$ &&25.92&-19.06 $\pm$ 0.15 &  $2.49_{-0.08}^{+0.09}$\\
46026 &41089 &3.605690 &-30.352664 & $10.86_{-8.30}^{+0.32}$ &&25.43&-19.92 $\pm$ 0.16 &  $1.47_{-0.03}^{+0.04}$\\
31763 &33358 &3.519867 &-30.366428 & $11.31_{-8.63}^{+0.20}$ &&26.92&-18.89 $\pm$ 0.17 &  $1.92_{-0.11}^{+0.11}$\\
42329 &38766 &3.513568 &-30.356804 & $11.83_{-7.93}^{+1.05}$ &$12.393_{-0.001}^{+0.004}$&26.88&-19.13 $\pm$ 0.18 &  $1.57_{-0.05}^{+0.06}$\\
46075&41179 &3.546722 &-30.352425 & $12.23_{-0.50}^{+1.38}$ &&26.90&-19.10 $\pm$ 0.19 &  $1.82_{-0.05}^{+0.11}$\\
70846 &53222 &3.498983 &-30.324758 & $12.50_{-0.15}^{+0.46}$ &&25.77&-20.79 $\pm$ 0.12 &  $1.27_{-0.02}^{+0.02}$\\ \hline

\end{tabular}\
\label{tab:sample}

\end{table*}

The final sample consists of 19 galaxies in the redshift range $9 < z < 12$. Among these four sources were included in the NIRSpec MSA (Micro Shutter Assembly) design of the UNCOVER spectroscopic follow-up \cite{fujimoto23,wang23}. The physical parameters of the galaxy candidates are listed in Table \ref{tab:sample} together with their spectroscopic confirmation $z_{\mathrm{spec}}$. Remarkably, we report a 100\% success rate in our spectroscopic confirmation, demonstrating the robustness of the photometric selection. Other studies \citep[e.g..][]{bunker_23b, hainline2023, haro23} also show high confirmation rate, 80\% or more.

\subsection{Simulated galaxies}\
The second step consists of computing the survey completeness function through the lensing cluster alongside its uncertainties. Following the procedure adopted in \citet{atek18}, we generated a set of 50,000 mock galaxies, which were randomly distributed directly in the source plane, which was constructed using the latest lensing model (see Section \ref{sec:Lensing}). The properties of these simulated objects were randomly allocated: the redshift in the range $ z= 8.5-12.5$, with intrinsic absolute magnitudes \muv\ from -23 to -15 mag. The input galaxy sizes follow a log-normal distribution and the size-luminosity relation derived for high-redshift galaxies, adopting the \cite{yang2022} determination for faint objects (\muv $> -17.2$) 
and \cite{shibuya15} for the bright sources.

A uniform distribution in the source plane, i.e. the physical plane, will undersample the regions of higher magnification in the image plane, leading to smaller statistics for the completeness function. For this reason, we decided to add a second layer of mock galaxies in the higher magnification area ($\mu >2$) within the source plane. 

Overall, this forward modeling naturally accounts for the galaxy shape distortion following the lensing shear, which directly affects the completeness function. It is very sensitive to the size, particularly at fainter magnitudes. Most importantly, the completeness function depends on several parameters, some of which are interdependent, such as the magnification and the source position, and need to be simultaneously accounted for. 

In order to compute the synthetic fluxes of the mock galaxies, we first generate SEDs models with {\tt BEAGLE} \citep{chevallard16} according to their simulated physical properties. The procedure uses stellar population models from \cite{Bruzual_Charlot_2003} and nebular emission models from \citep{gutkin16}. These templates, characterized by delayed star-formation histories $\psi(t)\propto t\exp(-t/\tau)$, and an SMC extinction law \citep{pei92}, and a constant metallicity Z = 0.1Z$_{\odot}$. Templates were redshifted and normalized to the observed magnitude within the F150W filter, corresponding to the rest-frame UV. In the end, we used {\tt GalSim} \citep{rowe2015galsim} to simulate the galaxies, which were in turn injected into the source plane of A2744, mapped into the image plane, and then inserted into the UNCOVER mosaics of A2744, placing 100 galaxies at a time. Then we follow the same procedure used for the observations to extract the sources and select dropout galaxies in the redshift range of $z = $ 9 - 12.

\section{The UV luminosity function at $z>9$} \
\label{sec:UVLF}
The observed galaxy number counts are computed in the uniform magnitude bins within the range of -22.5 $\leqslant$ M$_{UV} \leqslant$ -17.5 mag, with the size bin of $\Delta$mag = 1.0, 
except for the faintest bin which has a size of $\Delta$mag = 1.5. 
In our attempt to compute the UV LF, we also need to estimate the effective survey volume probed at each intrinsic magnitude (see Figure \ref{fig:LF_computation}). 

\begin{figure}
    \centering
    \includegraphics[width = 1.0 \linewidth]{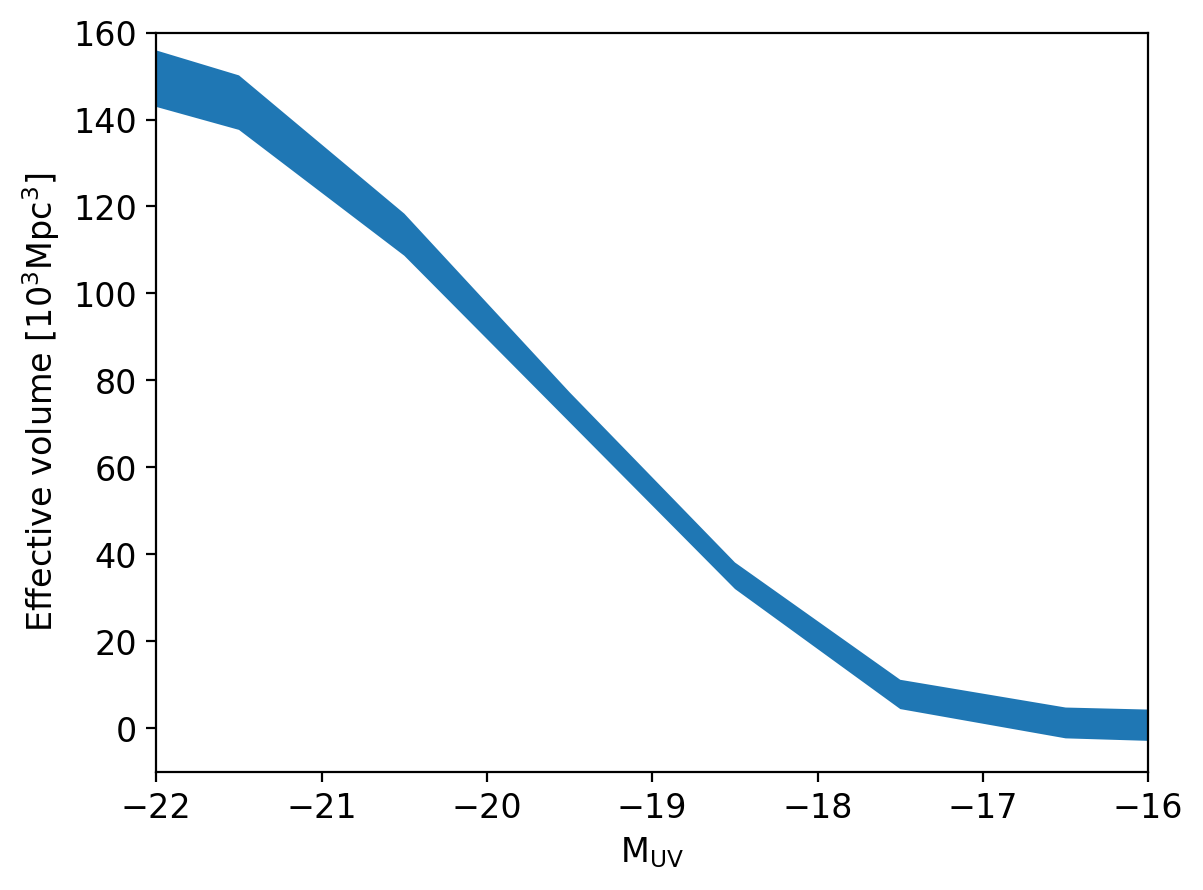}
    \caption{The effective volume as a function of the intrinsic absolute magnitude. The volume is computed from the completeness function and the surface area of the UNCOVER field. The 1$\sigma$ uncertainties are represented by the colored area around the curve.}
    \label{fig:eff_vol}
\end{figure}

To compute the survey volume, we first determine the survey area as a function of magnification and the redshift selection function. The effective volume consists of the co-moving maximal volume, which is distorted by gravitational lensing multiplied by the completeness function. As a result, the effective survey volume for each galaxy is expressed as:
\begin{equation}
    V_{eff, M_{UV, int}}=\int_{z=9}^{z=12}\int_{\mu=1}^{\mu=\mu_{max}} C(M_{UV,int},z)\frac{dV(\mu,z)}{dz}d\mu dz
\end{equation}
where, $C$ represents the completeness function, which is computed by comparing the output catalog with the original input one as a function of the intrinsic magnitude ($M_{UV, int}$), 
$\mu$ -- the magnification factor at the given redshift and $\mu_{max}$ is the magnification value at which a galaxy with a magnitude $M_{UV}$ can be detected, and $dV(\mu,z)$ represents the volume element available for the selection of a galaxy at a given redshift and amplification factor. The maximum volume depends mainly on the total surface area probed by the cluster, which is estimated to be $\sim 35$ arcmin$^{2}$. The completeness function values vary between 85\% on the bright-end and 10\% on the faint-end. Using this effective survey volume, the intrinsic UV LF can be calculated as follows:
\begin{equation}
    \centering
        \phi(M_i)dM_i=\frac{N_{obj,i}}{V_{eff}(M_i)}
\end{equation}
where $N_{obj, i}$ is the number of galaxies within each magnitude bin, and $V_{eff}(M_i)$ represents the effective survey volume corresponding to the $i$th bin of absolute magnitude $M_i$. The effective survey volume of the Abell 2744 cluster is presented in Figure \ref{fig:eff_vol}. We can see that the curve drops quickly at the bright-end, where the maximum volume depends mainly on the surface area with no magnification.

\begin{figure*}
\centering
\includegraphics[width=0.48\linewidth]{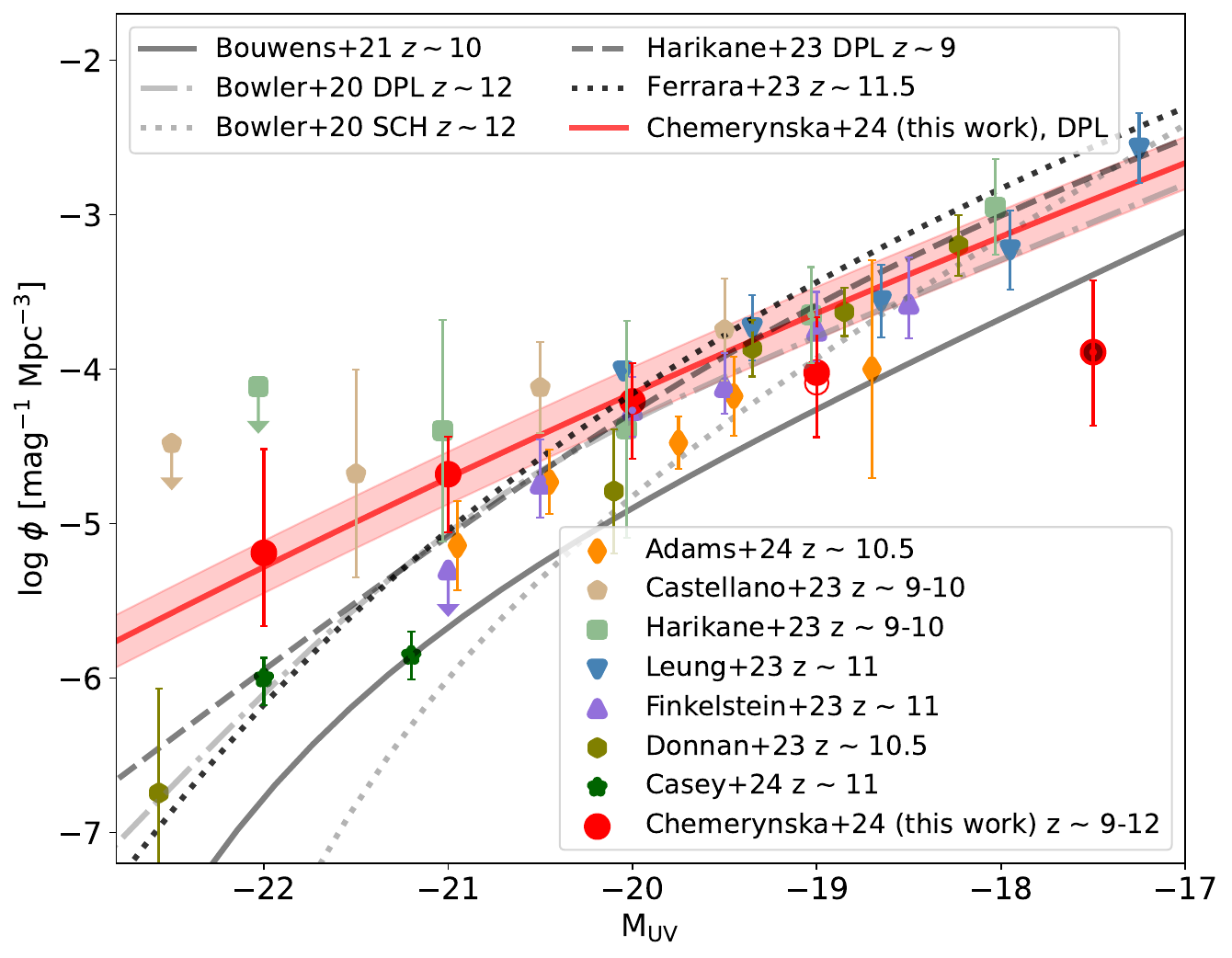}
\includegraphics[width=0.48\linewidth]{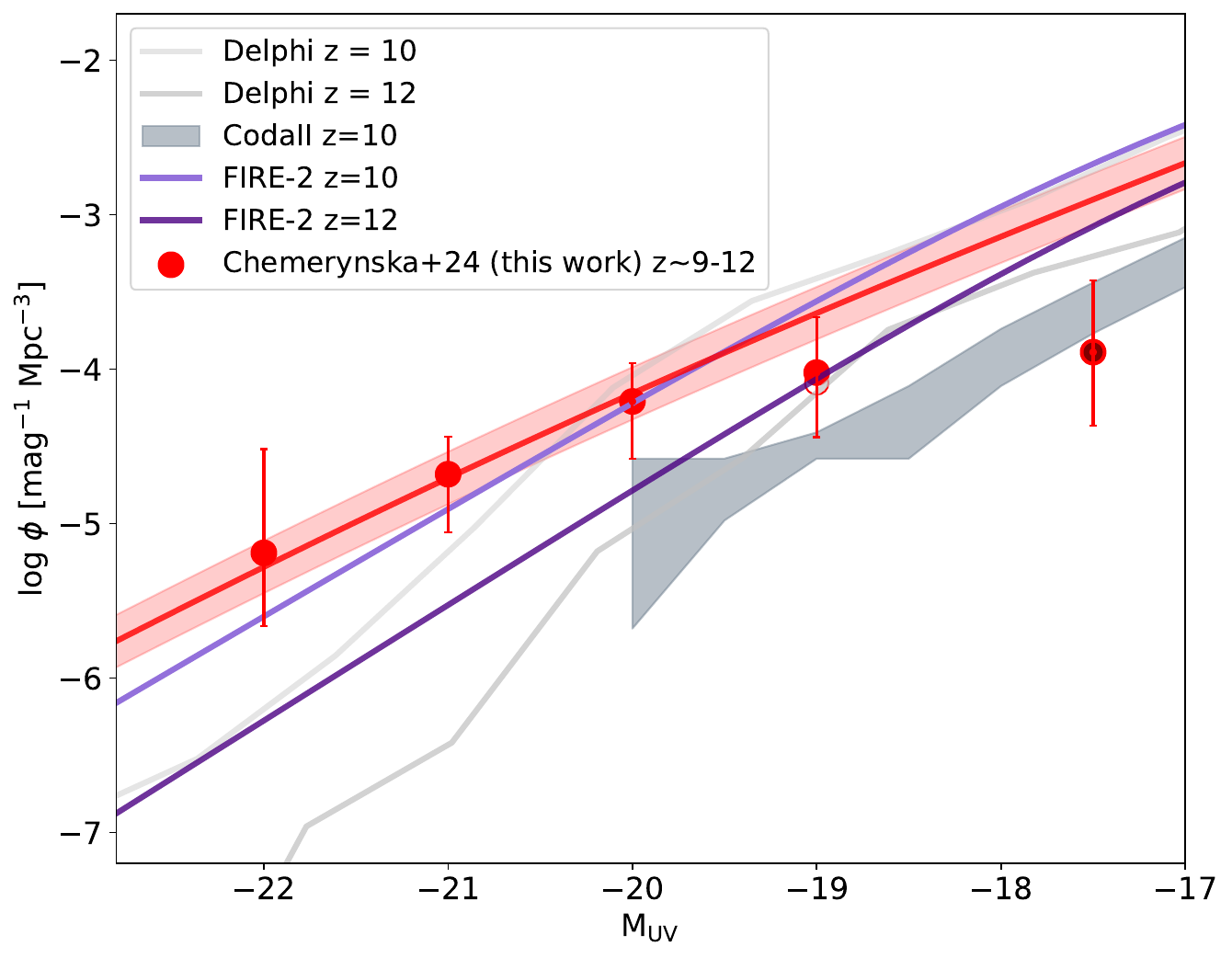}
\caption{The galaxy UV luminosity function at $9<z<12$ from the UNCOVER survey. {\bf Left:} The red circles, and associated error bars, represent our UV LF determination (the dark circle denotes our faintest bin, where the completeness is $\sim$10\%). The other points represent binned LF results from the literature: Recent results by \citet[][orange diamonds]{Adams23b}, \citet[][gold stars]{castellano23}, \citet[][green squares]{harikane23}, \citet[][blue triangles]{leung23}, \citet[][purple triangles]{finkelstein23}, \citet[][brown octagons]{donnan23} and \citet[][green stars]{Casey23}. Our best-fit DPL function is shown with a red solid line. {\bf Right:} Our UV LF results (red points and solid line) together with theoretical predictions. The light- and dark-grey solid lines represent the Delphi model estimates at z $\sim$ 10 - 12 \citep{mauerhofer23}, and the dark grey region corresponds to CoDaII constraints at $z = 10$ \citep{ocvirk20}. Also, recent results from FIRE-2 \citep{sun2023} at different redshifts are illustrated by solid light- and dark-purple lines. The red shaded area on both figures indicates uncertainties in the fitting.}
\label{fig:UV_mock+data}
\end{figure*}

To estimate the uncertainties of the UV LF, we considered several sources such as small-number Poisson statistics \citep{Gehrels86} 
and cosmic variance ($\sim$25-32\%), which was calculated following \cite{Trapp20}.  More importantly, we took into account statistical uncertainties in the lensing model, which impact both the amplification factor, the survey area, and to some extent the completeness. To estimate the systematic uncertainties of the lensing models, we compared two independent models constructed by \citet{furtak23a} and \citet{bergamini23b}. Overall, we did not find a significant difference between the two models, as the intrinsic uncertainties were in good agreement. Magnification uncertainties from the first lensing model were incorporated into the observed magnitude uncertainties in addition to the photometric errors. Then we used the Markov chain Monte Carlo (MCMC) simulations to explore the full error space for each galaxy, encompassing both photometric scatter and magnification factor uncertainties. For each of these iterations, we construct a luminosity function, which allows a galaxy to switch magnitude bins, hence changing the observed number counts. Regarding the effective survey volume, we include for each galaxy the $2-\sigma$ survey area uncertainty as a function of magnification. All these uncertainties were included in the final UV LF, incorporating the completeness errors in the process. 

\begin{table}
    \centering
    \caption[c]{Binned ultraviolet luminosity function at 9 $< z <$ 12. }
        \small{
            \begin{tabular}{lcc}
	       \hline
              \hline
	    M$_{UV}$ & N$_{obj}$  & $\log (\phi)$\\
                 &            &  [mag$^{-1}$Mpc$^{-3}$]\\
	       \hline
              \hline
              -22 & 1 & -5.19 $^{+0.67}_{-0.47}$\\
              -21 & 3 & -4.68 $^{+0.24}_{-0.38}$\\
              -20 & 7 & -4.21 $^{+0.25}_{-0.37}$\\
              -19 & 7 & -4.02 $^{+0.36}_{-0.42}$\\
              -17.5 & 1 & -3.89 $^{+0.46}_{-0.47}$\\
	   \hline
        \end{tabular}}	
\label{tab:UV LF}
\end{table}

Our determination of the UV LF is shown in Figure \ref{fig:UV_mock+data}, together with the recent results from other surveys in the same redshift range. The tabulated values of UV LF are presented in Table \ref{tab:UV LF}. In general, our sample consists of relatively bright galaxies (\muv\ < -19), except for one faint galaxy at (\muv\ < -17.5). The most important result of our study is the apparent overabundance of bright galaxies compared to theoretical expectations \citep{mason15, Tacchella18, Harikane22, mauerhofer23} or extrapolations from lower-redshift determinations. This excess is in broad agreement with the most recent \jwst\ results that show a similar trend \citep{harikane23,bouwens23,finkelstein22,McLeod23,adams23}. 

We fit our LF data points with a double power-law function \citep[e.g.,][]{bowler20,finkelstein2210,donnan23}, which better describes the overall functional form of the LF at high redshifts compared to a classical Schechter \citep{schechter76} function.
\begin{equation}
    \phi(M) = \dfrac{\phi^*}{10^{0.4(M-M^*)(\alpha+1)}+10^{0.4(M-M^*)(\beta+1)}}
\end{equation}
where $\phi^*$ is a normalization, $M^*$ represents characteristic magnitude, $\alpha$ and $\beta$ are the faint-end slope and the bright-end slope, respectively. Because we mainly probe the bright-end of the LF, we chose to combine robust literature data points \citep{harikane23,leung23,donnan23} and for the faint-end with our results on the bright-end (\muv < -19 mag). To determine the best-fit function, we applied the Levenberg-Marquardt (LM) approach using MCMC simulations. We fixed the parameters 
$M^{*}$ = -20.67 and $\alpha$ = -2.1 in the fitting procedure, which is similar to the value adopted by recent \jwst\ studies at similar redshift \citep{harikane23,adams23}. 

The results are presented in Table \ref{tab:sfrd}, together with the results of recent studies. The best-fit DPL function is represented by the red curve in Figure \ref{fig:UV_mock+data}. We can see that the DPL fits reasonably well the bright-end of the LF, while our determinations at the faint-end (which are not included in the fit) are lower, with larger uncertainties. We find a shallower bright-end slope ($\beta=-2.66\pm1.09$) compared to other studies, albeit with large uncertainties. Our results are in better agreement with \citep{castellano23}, whose galaxy sample has a significant overlap with our study, although they used a simple treatment of lensing effects. We also compare our results with the UV LF predicted by galaxy formation models in a similar redshift range. Our results are above most theoretical determinations from hydrodynamical simulations \citep{ocvirk20,sun2023} and semi-analytical models \citep{mauerhofer23}. We note that the FIRE (Feedback In Realistic Environments) simulations reproduce better our results at $z=10$.

\section{The star formation rate density}
\begin{figure*}
    \centering
    \includegraphics[width = 0.8 \linewidth]{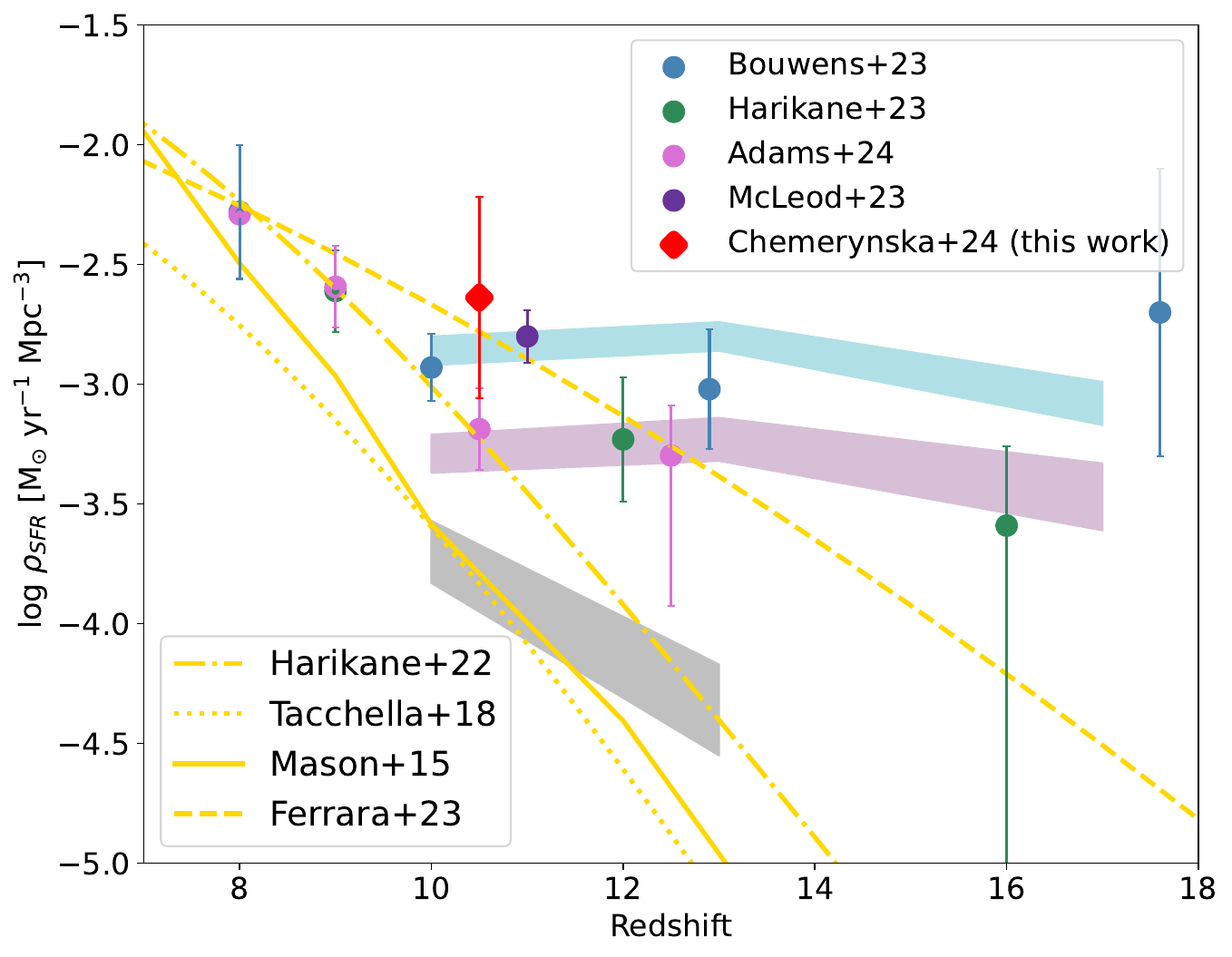}
    \caption{The star formation rate density (SFRD) at $z\sim10.5$. Our SFRD result, integrated down to -17 mag, is shown by a red circle. We also show a selection of literature results covering a similar redshift range: \citet[][blue points]{bouwens23}, \citet[][green points]{harikane23}, \citet[][pink points]{Adams23b} and \citet[][purple points]{McLeod23}. The SFR densities derived by \citet{bouwens22b} at $z \sim 10-17$ are shown in light grey, dark pink, and blue shaded regions, which correspond in their definition to ``robust'', ``solid'', and ``possible'', candidates, respectively. The gold lines represent theoretical predictions for the cosmic SFR density as a function of redshift based on the models of \citet[][solid line]{mason15}, \citet[][dotted line]{Tacchella18}, and \citet[][dashed line]{Ferrara23}.
    The yellow dot-dashed line is the best-fit result of \citet[]{Harikane22} at $z<7$, which was extrapolated to higher redshifts ($z>10$) assuming a constant star formation efficiency.
    }
    \label{fig:SFRD}
\end{figure*}

\
\label{sec:sfrd}
We compute the UV luminosity density by integrating the best-fit DPL of luminosity function down to a magnitude limit of \muv\ = - 17 mag, following recent literature results. Adopting a fixed faint-end slope $\alpha$ = -2.1, we obtain log($\rho_{UV}$/ergs s$^{-1}$ Hz$^{-1}$ Mpc$^{-3}$) = 25.3, which is slightly higher than recent \jwst\ results. A full comparison with recent literature results is presented in Table \ref{tab:sfrd}, alongside the best-fit luminosity function parameters. Furthermore, we derive the star-formation rate density ($\rho_{SFR}$) using a canonical conversion factor $K_{UV}$ = 1.15$\times$10$^{-28}$ M$_{\odot}$ yr$^{-1}$/erg s$^{-1}$ Hz$^{-1}$ \citep{madau2014}. In Figure \ref{fig:SFRD}, we show how our results compare with literature values and galaxy formation models. At similar redshifts, our value is slightly higher than values derived from the main \jwst\ surveys \citep{adams23,bouwens23,harikane23, McLeod23}. Within the uncertainties, our result is closer to the results of \citet{McLeod23} and \citet{bouwens23}, when using their entire sample, including ``possible'' high$-z$ candidates. 

From the theoretical viewpoint, our results also appear at odds with galaxy formation models. There has been a long-standing debate about the redshift-evolution of the star formation efficiency (SFE) during the pre-\jwst\ era. At redshifts higher than $z=9$ several models argued for a constant SFE, which were able to reproduce \hst\ observations \citep[e.g.][]{mason15,oesch18}. In Figure \ref{fig:SFRD} we compare our results with predictions of a few examples of such models. Our derived SFRD value is 4 to 10 times higher than values predicted by constant-SFE models \citep{mason15, Tacchella18, Harikane22}. Together with literature results at other redshifts, this shows a clear excess relative to predictions of constant-SFE models. Given the reliability of our photometric selection, sample contamination from low-redshift sources is highly unlikely to explain this excess. Additionally, our results are in agreement with the prediction by \cite{Ferrara23} for radiation-driven outflows that clear dust. According to this model, at redshift $z\sim$10, the cosmic specific star formation rate (sSFR) reaches a critical value, leading to almost half of the galaxies showing super-Eddington driving outflows. The model curve is in good agreement with the observed abundance of early bright galaxies.

In a recent spectroscopic follow-up with NIRSpec, \citet{fujimoto23} has explored possible AGN natures for 4 spec-$z$ confirmed sources at $z\geq8.5$ in our sample. One candidate (ID 26185) is detected in the observed-frame 2–7 keV band in Chandra observations at a $\sim 4 \sigma$ level \citep{bogdan23,goulding23}. Another candidate (ID 20466) shows a broad feature in the H$\beta$ emission line profile but it is not part of our sample. In order to assess the contribution of these potential AGN sources to the UV LF, and to what extent they can explain the overabundance of bright galaxies at $z>9$, we recompute the UV LF
without the AGN source. The result is shown with the red empty circle in Figure \ref{fig:UV_mock+data}. The main results remain unchanged. The slightly lower value at the \muv\ = -19 mag bin does not affect the best-fit DPL solution. Therefore, the estimated SFRD in Figure \ref{fig:SFRD} remains virtually the same. 

Many recent studies hint at the existence of a substantial population of AGN at $5<z<8$ \citep{matthee23,larson23,kokorev23}, with a redder rest-frame optical than UV emission, and which appear more numerous than UV-selected AGN. \citep{labbe23b,furtak23c,greene23}. Other AGNs have also been identified at $5<z<7$, based on their broad H$\alpha$ emission line \citep{harikane23b,maiolino23}. The exact contribution of AGN to the UV luminosity function remains unclear at this point, as their UV emission could originate from star formation as well as dust-scatted AGN light. Further SED modeling, observations in the infrared, and spectroscopic line diagnostics will be needed to obtain stronger constraints on the AGN contribution.   

Other explanations for the UV LF enhancement at high redshift involve an evolution of the physical properties of galaxies and the star formation processes. A decrease in the dust attenuation at higher redshift as a result of strong outflows expelling gas and dust out of these early galaxies has been suggested in \citet{ferrara22}. However, this strong redshift-evolution of the dust attenuation is currently not supported by the UV colors of galaxies across $6<z<12$ \citep{finkelstein23}. Alternatively, a modification of the stellar mass-to-light ratio can also explain the UV enhancement while the underlying mass function remains unchanged. One of the physical processes that can cause such variation is a stochastic star formation history. Rapid SFR variations can lead to significant deviations in the M/L ratio, making the UV luminosity an unreliable tracer of the stellar mass function. However, recent theoretical efforts have shown significant disagreements regarding the level of this UV luminosity variation. A model including variability in the physical processes (SFR, dust, metallicity) predicts the scatter of lower mass galaxies to brighter luminosity bins due to the steep slope of the LF. A UV luminosity scatter of $\sigma_{\rm UV}=1.75$ is sufficient to reproduce the observed UV excess \citep{shen23,sun2023}. On the other hand, hydrodynamical simulations that include stochasticity result in significantly lower values around $\sigma_{\rm UV}=0.6$, which remain below the levels required to match observations \citep{pallottini23}. Using SED modeling that includes stochastic SFHs, \citet{ciesla23} reach similar conclusions.
\cite{pacucci22} show that AGN at high redshifts can also contribute to such UV excess.
Finally, another possible cause for the M/L deviations is a different initial mass function (IMF) at early epochs. A different IMF, with a characteristic mass of $M_{c}=10$ M$_{\odot}$, would decrease the M/L ratio by a factor of several \citep{raiter10}. Although changes in the IMF are expected at early epochs \citep[e.g.;][]{steinhardt23, woodrum23, wang23a}, no observational evidence of such an evolution has been reported yet.

\begin{table*}
    \centering
        \caption{The best-fit parameters for the double power-law function of recent studies together alongside luminosity consistency and star formation density. Values presented with an asterisk indicate the parameter is fixed to this value in the fitting procedure.}
    \begin{tabular}{lcccccccc}
    \hline
    \hline
         Study&  z& $\log (\phi)$  & $M^{*}$  & $\alpha$ & $\beta$ & $\log (\rho_{UV})$ & $\log (\rho_{SFR})$ & K$_{UV}$  \\
         &  & [mag$^{-1}$Mpc$^{-3}$]  & [mag]  & & & [ergs s$^{-1}$ Hz$^{-1}$ Mpc$^{-3}$ ] & [M$_{\odot}$yr$^{-1}$Mpc$^{-3}$] &  [M$_{\odot}$ yr$^{-1}$erg$^{-1}$ s Hz]\\
    \hline
    \hline
         Harikane+23&  9&  -3.50$^{+1.53}_{-0.65}$& -19.33$^{+2.24}_{-0.96}$ & -2.1$^*$ & -3.27$^{+0.34}_{-0.37}$ & 25.28$^{+0.19}_{-0.16}$ & -2.61$^{+0.19}_{-0.16}$ & 1.15$\times$10$^{-28}$  \\
         Bouwens+23&  10& -3.55$^{+0.17}_{-0.12}$  & -19.67$^*$ & -2.35$^*$ & -3.75$^*$ & 25.22 
         $\pm$ 0.14 & -2.93 
         $\pm$ 0.14 & 0.7$\times$10$^{-29}$  \\
         Adams+24&  10.5& -5.02$^{+0.47}_{-0.39}$ & -21.10 $^{+0.78}_{-0.64}$ & -2.1$^*$ & -4.45 $^{+0.97}_{-1.02}$ & 24.75
         $\pm$ 0.17 & 3.19
         $\pm$ 0.17 & 1.15$\times$10$^{-28}$  \\
         Leung+23&  11& -4.78$^{+0.15}_{-0.16}$ & -20.99$^*$ & -2.22$^{+0.23}_{-0.23}$ & -4.19$^*$ &  &  &   \\
         McLeod+23&  11  &-4.69 $\pm$ 0.45 & -20.87 $\pm$ 0.63 & -2.35$^*$ & -4.16 $\pm$ 0.76 & 25.17$^{+0.11}_{-0.11}$ & -2.77$^{+0.11}_{-0.11}$ & 1.15$\times$10$^{-28}$  \\
         This work&  10.5& -4.22 $\pm$ 0.71 & -20.67$^*$ & -2.1$^*$ & -2.66 $\pm$ 1.04 & 25.30 $\pm$ 0.42 & -2.64 $\pm$ 0.42  & 1.15$\times$10$^{-28}$  \\

    \hline
    \end{tabular}
    \label{tab:sfrd}
\end{table*}

\section{Conclusions}\
\label{sec:Conclutions}

In the present paper, we used a sample of $z>9$ galaxies discovered in the UNCOVER survey to construct a robust UV luminosity function across $9<z<12$. This high-redshift catalog consists of 19 galaxies observed behind the lensing cluster A2744 with intrinsic magnitudes between M$_{UV}\sim$ -22 and -17 mag. Sources were selected using a combination of color-color dropout criteria and photometric redshift estimates derived from SED fitting \citep{atek23}. Among these sources, four were followed-up with NIRSpec Prism observations as part of the UNCOVER survey. All four sources were confirmed around their photometric redshift \citep{fujimoto23}, achieving a 100\% confirmation rate, which underscores the reliability of the present photometric sample.

To compute the most accurate UV luminosity function at $z>9$ through lensing clusters, we first conducted a forward modeling procedure to derive the effective survey volume. By operating directly in the source plane, we naturally take into account all lensing effects. Furthermore, in order to assess systematic uncertainties related to lensing, we compared two independent models derived by \citet{furtak23a} and \citet{bergamini23b}. 

Despite, a careful estimate of the cosmic variance and its inclusion in our UV LF procedure, uncertainties main remain, particularly between different fields and methods. For instance, our determination ranges from 25\% to 32\%, while \citep{castellano23} values are in the range 13-15\%, despite probing a slightly smaller survey area. In view of the numerous independent estimates of the UV LF at these redshifts, it is reassuring to see that most results agree on the overabundance of UV-bright galaxies  \citep[cf. also][]{CANUCS}. For instance, a large compilation of \jwst\ deep fields has shown a steeper bright-end, consistent with individual results \citep{harikane23b,McLeod23,Adams23b}. \citet{finkelstein23} also shows that, despite the significant cosmic variance, the UV LF determinations are still higher than most theoretical models. Overall, it is clear that both cosmic variance and small number counts are the major uncertainties on the UV LF bright-end. The way-forward is to advocate for more \jwst\ Cosmos-Web-type of surveys, and upcoming observations from the {\em Euclid} wide surveys. 

The most important result is the overabundance of UV-bright galaxies, which reach a factor of 10-100 higher than theoretical expectations, or previous \hst\ observations. Our results confirm the emerging picture describing a strong redshift-evolution of the physical properties or the physical processes governing star formation in early galaxies \citep{harikane23,finkelstein23,bouwens23}. At the same time, given the depth of these lensing-assisted observations and the supposedly steep faint-end slope of the LF at these redshifts, there is an apparent lack of faint galaxies around \muv>-17 mag. Whether this is due to cosmic variance or the effects of small completeness values that reach $\sim 10\%$ or less, remain to be determined. Deeper imaging programs such as GLIMPSE (PI: Atek, PID 3293) will be able to confidently constrain this very faint end of the UV LF.

Our determination of the star formation rate density at $z\sim10.5$ lies above most theoretical models of galaxy formation \citep{mason15,ocvirk20,mauerhofer23}. Among the variety of scenarios proposed to explain this excess, star formation stochasticity has been widely explored. For instance, by including variability in the conversion from dark matter halo mass to UV luminosity, a dispersion in the UV luminosity of $\sigma_{\rm UV}$ = 1.5 to 2 is required to match the observations at these redshifts \citep[e.g.][]{shen23}. At the same time, other theoretical efforts based on hydrodynamical simulations or stellar population modeling have measured an insufficient variability to explain such excess \citep{pallottini23,ciesla23}. While early studies indicate that high$-z$ galaxies experience bursty star formation \citep{endsley23,looser23}, spectroscopic observations of a large sample of these sources will help constrain the stochastic SFH and the inferred UV luminosity dispersion. A combination of SED modeling associated with observational constraints of SFR indicators on different timescales is key to addressing this question. Such efforts will be largely complemented by wide-area surveys, such as {\em Euclid}, which will uncover thousands of galaxies brighter than \muv\ = -21 mag at redshifts higher than $z=8$. Large ground-based spectroscopic follow-up campaigns will be essential for their redshift confirmation.   

Low-mass galaxies are predominantly affected by stellar feedback, which causes significant variations in their star formation history. Therefore, they provide the best sites to study the strength and effects of stochasticity on the observed physical properties in general. 
Additionally, by observing the well-studied Abell 2744 field with all available medium-band NIRCam filters, it is possible to identify and characterise various sources that emit strong signals from the cluster through the era of reionization (PI: Suess, PID: 4111). Also, the \jwst-GO-3516 program (PI: Matthee) proposed a NIRCam grism survey around the powerful lensing cluster Abell 2744 to detect emission lines at high redshifts to identify the most metal-poor pockets of star formation and measure the ionizing photon production efficiency of these dwarf galaxies.  
Ultra-deep \jwst\ imaging surveys like GLIMPSE are set to measure the prevalence of ultra-faint galaxies at $z>6$ and provide constraints on the nature and strength of feedback in galaxy formation models. Programs from \jwst\ Cycle 1, such as CANUCS (PI: Willott, PID: 1208), PEARLS (PI: Windhorst, PID: 1176) and more, are going to provide many more fields, albeit at shallower depths to increase the sample size of these lensed high-redshift galaxies.

Such surveys will also probe the so far elusive, yet supposedly common, population of faint galaxies during the Dark Ages, to test whether this surprising redshift-evolution of bright galaxies also applies to low-mass galaxies, which are expected to be the dominant population at early epochs.

\section*{Acknowledgments}\
This work is based on observations obtained with the NASA/ESA/CSA \textit{JWST} and the NASA/ESA \textit{Hubble Space Telescope} (HST), retrieved from the \texttt{Mikulski Archive for Space Telescopes} (\texttt{MAST}) at the \textit{Space Telescope Science Institute} (STScI). STScI is operated by the Association of Universities for Research in Astronomy, Inc. under NASA contract NAS 5-26555. This work was supported by CNES, focused on the \jwst\ mission. This work was supported by the Programme National Cosmology and Galaxies (PNCG) of CNRS/INSU with INP and IN2P3, co-funded by CEA and CNES. This work has made use of the \texttt{CANDIDE} Cluster at the \textit{Institut d'Astrophysique de Paris} (IAP), made possible by grants from the PNCG and the region of Île de France through the program DIM-ACAV+, and the Cosmic Dawn Center and maintained by S. Rouberol. AZ acknowledges support by grant No. 2020750 from the United States-Israel Binational Science Foundation (BSF) and grant No. 2109066 from the United States National Science Foundation (NSF); by the Israel Science Foundation grant No. 864/23; and by the Ministry of Science \& Technology, Israel. PD acknowledges support from the NWO grant 016.VIDI.189.162 (``ODIN") and from the European Commission's and University of Groningen's CO-FUND Rosalind Franklin program. 

\section*{Data Availability}
The data underlying this article are publicly available on the \texttt{Mikulski Archive for Space Telescopes}\footnote{\url{https://archive.stsci.edu/}} (\texttt{MAST}), under program ID 2561. Reduced and calibrated mosaics, as well as lensing products, are available on the UNCOVER webpage: \url{https://jwst-uncover.github.io/}



\bibliographystyle{mnras}
\bibliography{reference} 





\appendix
\section{The impact of binning parameters on the UV LF}

\begin{figure}
    \centering
    \includegraphics[width = 1.0 \linewidth]{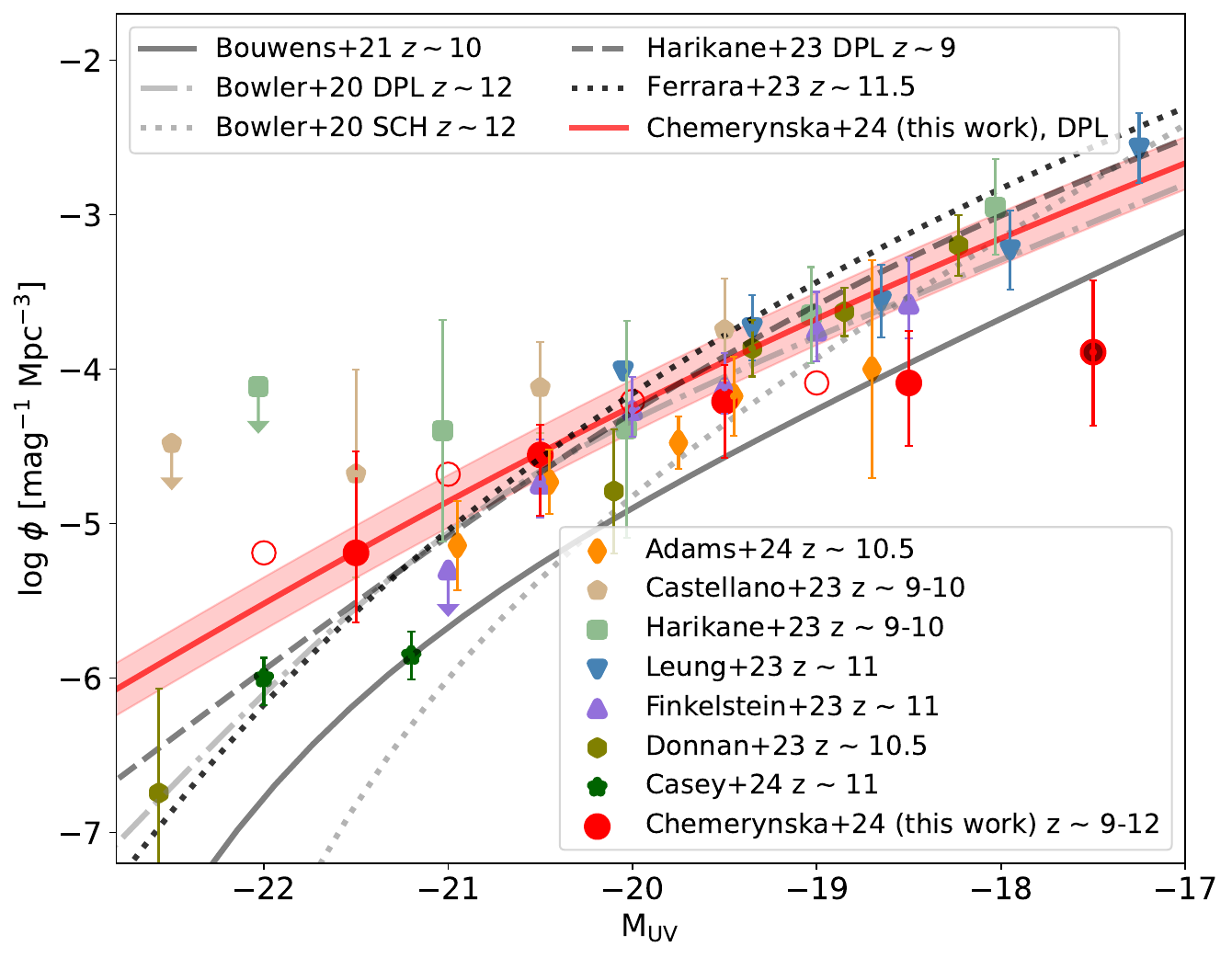}
    \caption{The galaxy UV luminosity function at $9<z<12$ from the UNCOVER survey. The empty circles represent the fiducial UV LF determination in this paper, whereas the red-filled circles represent the UV LF resulting from a different binning scheme. The rest of the legend and references are the same as in Figure \ref{fig:UV_mock+data}.}
\label{fig:UV_a}
\end{figure}

When computing the UV LF, the choice of the binning scheme, including the position and the width of the bins, can vary. While we used bins starting at -22, we explored the effects of adopting different binning boundaries. We recalculated the outcomes with a shift of 0.5 magnitudes towards fainter values. The galaxy counts in the new bins become [1,4,7,6,1]. The re-binned UV LF is presented in Figure \ref{fig:UV_a}.
The two brightest bins move slightly downwards, more in line with \citet{harikane23b} and still significantly higher than the predictions. The best-fit parameters for the double power low are $M^*$ = 19.97 mag, $\alpha$ = -2.1, $\beta$ = -2.85 $\pm$ 1.00, and $\log(\phi)$ = -3.92 $\pm$ 0.72 [mag$^{-1}$Mpc$^{-3}$].

\label{lastpage}
\end{document}